\documentclass[prl,aps,preprint,showpacs,floatfix]{revtex4}
\usepackage{graphicx}

\begin{document}
\title{Parameter scaling in the decoherent quantum-classical \\ 
transition for chaotic systems}
\author{Arjendu K. Pattanayak$^{(a)}$ and Bala Sundaram$^{(b)}$}
\date{\today}
\affiliation{(a) Department of Physics, Carleton College, Northfield, 
Minnesota 55057\\ (b) Graduate Faculty in Physics \& Department of 
Mathematics, \\ CSI-CUNY, Staten Island, New York 10314}

\begin{abstract}
{The quantum to classical transition has been shown to depend on a number 
of parameters. Key among these are a scale length for the action, $\hbar$, 
a measure of the coupling between a system and its environment, $D$, and, 
for chaotic systems, the classical Lyapunov exponent, $\lambda$. We
propose computing a measure, reflecting the proximity of quantum and
classical evolutions, as a multivariate function of $(\hbar,\lambda,D)$ 
and searching for transformations that collapse this hyper-surface 
into a function of a composite parameter $\zeta =
\hbar^{\alpha}\lambda^{\beta}D^{\gamma}$. We report results
for the quantum Cat Map, showing extremely accurate scaling behavior 
over a wide range of parameters and suggest that, in general, the
technique may be effective in constructing universality classes in
this transition. }
\end{abstract}

\pacs{PACS numbers: 05.45.Mt,03.65.Sq,03.65.Bz,65.50.+m}
\maketitle

The classical description of a system approximates the inherently quantum 
world and has significantly different predictions. The question of when 
quantum mechanics reduces to classical behavior is both fundamentally
interesting as well as relevant to applications such as quantum computing 
which seek to exploit this difference. The quantum to classical transition 
(QCT) is now understood to be affected not only by the relative size 
of $\hbar$ (Planck's constant) for a given system, but also by $D$, a measure 
of the coupling of the environment to the quantum system of interest, an 
effect termed decoherence. Further, in systems where the classical evolution 
is chaotic, the transition is also affected by the chaos in the system, and 
thus by $\lambda$, the Lyapunov exponent of the classical trajectory 
dynamics~\cite{zp,pb}. As such, the QCT for chaotic Hamiltonians is, in 
general, a complicated function of multiple parameters, and is far from 
being fully understood.

However, the parametric dependence is as daunting as it first appears, 
particularly near the transition regime. Several studies point to 
composite parameters, indicating that the transition is not independently 
affected by each of the three parameters. For example, 
considerations~\cite{ott,doron,zp,koslovsky,pb,ap} of stochastic
quantum evolution or a master equation show that the parameter range for 
classical behavior is not simply $\hbar \ll 1$ but depends also on
$D$. These and similar studies also indicate scaling relationships 
involving $\hbar,D,\lambda$. Other work has addressed correspondence at 
level of trajectories which requires a continuous extraction of information
from the environment~\cite{tanmoy}, as opposed to tracing over these
variables. However, here again, the condition for correspondence may be 
viewed as a composite variable where $D$ is appropriately replaced by the 
strength of the measurement. More recently, it has been argued that 
Hamiltonian systems fall into a range of universality classes with 
distinctly different QCTs~\cite{salman}, behavior 
manifested in the density matrix far from the transition regime. 

With these as motivation, we propose that significant progress can be made
by (a) computing measures which directly reflect the `distance' between 
quantum and classical evolutions as a function of  
$\hbar,\lambda,\; \mbox{and}\;D$ and then (b) searching for transformations 
that collapse the resulting hyper-surface onto a function of a composite 
parameter of the form $\zeta = \hbar^{\alpha}\lambda^{\beta}D^{\gamma}$. 
The aims are (i) to search for this scaling, especially the coefficients 
$\alpha,\beta,\gamma$~\cite{fn1}; (ii) to investigate the range 
of parameters and initial conditions over which the scaling holds and 
(iii) to study the dependence of the distance measure on $\zeta$. 
We can anticipate the possible outcomes:  First, that $\alpha,\beta,\gamma$ 
are independent of the Hamiltonian. If this extremely unlikely scenario
holds, we have a modified Planck's constant governing all quantum 
chaotic systems, and universality classes are differentiated by differing 
dependences of the distance measure on $\zeta$. Second, a range of behavior 
for $\alpha,\beta,\gamma$ is seen, including a dependence on initial 
conditions, providing a classification scheme possibly correlated with 
the previously proposed classes. Finally, any scaling may be associated 
with the nature (single-scale, multi-scale) of the quantum coherence 
affected by the environment. This suggests a third alternative where scaling 
behavior exists only for limited classes of systems or limited parameter 
ranges, in which case the existence or range of scaling defines universality 
classes. 

Below, we present broad arguments for the existence of such scaling.
We then consider two alternate measures of the quantum-classical 
distance including a generalized Kullback distance~\cite{schlogl}. 
We numerically test our ideas with these measures on a specific system, 
the noisy quantum Cat Map. For the Cat Map, the Lyapunov exponent is 
a constant, such that the QCT is at most a two-parameter transition. 
We show that this two-parameter transition, in fact, reduces to an 
effective single-parameter transition. This scaling is remarkably 
sharp and extends over a large range of parameters. In the case of 
the Cat Map, the quantum nature of the system is a well-defined 
function of $\zeta\equiv \hbar^2\lambda D^{-1}$, consistent with previous
analysis~\cite{koslovsky}. We discuss the nature of the transition 
in some detail, and conclude with expectations for the decoherent 
QCT in other, more general, chaotic systems.

We begin from the equation describing the evolution of a quantum Wigner 
quasi-probability $\rho^W$ under Hamiltonian flow with potential $V(q)$ 
while coupled to an external environment~\cite{zp}:
\begin{eqnarray}
{\partial \rho^W\over\partial t}=\{H,\rho^W\}
&+&\sum_{n \geq 1}\frac{\hbar^{2n}(-1)^n}{2^{2n} (2n +1)!}
\frac{\partial^{2n+1} V(q)}{\partial q^{2n+1}}\;
\frac{\partial^{2n+1} \rho^W}{\partial p^{2n+1}}\nonumber \\
&+& D \nabla^2\rho^W.
\label{wigner}
\end{eqnarray}
The first term on the right is the Poisson bracket, generating the
classical evolution for $\rho^W$. The terms in $\hbar$ add the quantal
evolution while the effects of the environmental coupling are reflected in
the diffusive term. For simplicity, we couple to all phase-space variables, 
although the results generalize. Consider for the moment only the 
classical evolution in the presence of the environmental perturbation. 
As a result of chaos, the density $\rho$ develops fine-scale structure 
exponentially rapidly, with a rate given by a generalized Lyapunov 
exponent.  When the structure gets to sufficiently fine scales, the 
noise becomes important. The basic role of noise is to wipe out, or 
coarse-grain, small-scale structure. The competition between chaos and 
noise leads to a metastable balance for the fine-scale 
structure~\cite{physicaD}. This is clearly visible in the measure 
$\chi^2 \equiv \frac{{\rm Tr}[\rho^W \nabla^2\rho^W]}{{\rm Tr}
[(\rho^W)^2]} = - \frac{{\rm Tr}[|\nabla \rho^W|^2]}{{\rm Tr}
[(\rho^W)^2]}$
where the second equality results from an integration by parts. 
This quantity $\chi^2$ is approximately the mean-square radius of
the Fourier expansion of $\rho$ and, for our purposes, measures the structure 
in the distribution~\cite{97_1}. For a classically chaotic system under 
the influence of noise, $\chi^2$ settles after a transient to the metastable 
value $\chi^{2*} = \sum_i \Lambda^+_{2,i}/2D \equiv \Lambda/2D$ where 
the $\Lambda^+_{2,i}$ are $\rho$ dependent versions of the usual 
generalized positive Lyapunov exponents of second 
order~\cite{schlogl,physicaD}.

Now let us add quantal corrections to the mix. As seen from 
Eq.~(\ref{wigner}), the terms are of the form
$\hbar^{2n} \frac{\partial^{2n+1} V(q)}{\partial q^{2n+1}}\;
\frac{\partial^{2n+1} \rho^W}{\partial p^{2n+1}}$ which
scale as $\hbar^{2n}\chi^{2n+1}V^{(2n+1)}(x)$, where 
$V^{(r)}$ denotes the $r$th derivative of $V$. Since $\chi^2$ 
settles to the fixed value $\Lambda/2D$, this contribution to the difference 
between the quantum and classical evolution may be estimated to be 
$\zeta\equiv\hbar^{2n} \Lambda^{n+1/2}D^{-(n+1/2)} V^{(2n+1)}(x)$ where
$x\approx \chi^{-1}=\sqrt{D/\Lambda}$. Therefore, quantum-classical 
distances should scale, in complete generality, with the single 
parameter $\zeta$ for small $\zeta$. The particular form of $\zeta$ is 
decided by the details of the Hamiltonian and, in general, the scaling 
relationship is deduced from a direct examination of the 
deviation of the quantal propagator from the classical version. 

As a measure of the distance between two distributions $P$ and $Q$ 
with support on the same space, we introduce the quantity
\begin{eqnarray}
K_\epsilon(P,Q) &=& \frac{1}{\epsilon}[\ln({\rm Tr}[P Q^\epsilon])
                 -\ln({\rm Tr}[P^{1+\epsilon}])\nonumber \\
                &+& \ln({\rm Tr}[P^\epsilon Q])
                 -\ln({\rm Tr}[Q^{1+\epsilon}])]
\label{eq:K}
\end{eqnarray}
where ${\rm Tr}$ denotes the trace over all variables. 
$K_\epsilon$ is a generalized Kullback-Liebler (K-L) distance, 
reducing to a symmetrized form of the usual K-L distance~\cite{schlogl} in the 
limit $\epsilon \to 0$. To see this, use that 
$P^\epsilon = \exp(\epsilon\ln P) \approx 1 + 
\epsilon\ln(P) + {\cal O}(\epsilon^2)$ for $\epsilon\to 0$. 
Then, the first term in Eq.~(\ref{eq:K}) becomes 
$\frac{1}{\epsilon}\ln({\rm Tr}[P Q^\epsilon])
\approx \frac{1}{\epsilon}\ln(1 + \epsilon {\rm Tr}[P\ln Q]).$
Now using the expansion for $\ln(1+x)$ for small
$x$ for this and the other terms, this yields
\begin{equation}
\lim_{\epsilon \to 0} K_\epsilon(P,Q) =
{\rm Tr}[P\ln(\frac{Q}{P})] + {\rm Tr}[Q\ln(\frac{P}{Q})]
\end{equation}
which is indeed a symmetrized version of the usual K-L distance.
$K_\epsilon$ has similar properties, and is a general measure 
of the distance between the two probability distributions. When $P$
and $Q$ are identical, this measure is zero. A 
convenient form of $K_\epsilon$ is for $\epsilon =1$ when 
it reduces to 
\begin{equation}
K_1(P,Q) = \ln \bigg[\frac
{({\rm Tr}[PQ])^2}{{\rm Tr}[P^2]{\rm Tr}[Q^2]}\bigg].
\end{equation}
We begin from an initial phase-space 
distribution $\rho_0$, which is propagated in time using separately 
(i) the quantum dynamics to yield $\rho_W(t)$ and (ii) the 
classical dynamics for $\rho_c(t)$. During the propagation, 
the distance $K_1(\rho_W,\rho_c)$ is monitored. The initial 
distance $K_1(t=0) =0$, and due to diffusive noise all initial 
distributions relax to the constant distribution, such that 
$K_1(t\to\infty)=0$ and hence $K_1$ is bounded as a function of 
time. For a given set of parameters $\hbar,D$ and for some 
reasonably long time $t_m (\gg 1/\Lambda)$, the maximal value 
of $K^m_1(\rho_W,\rho_c)$ is our measure of the 
quantum-classical distance. 

\begin{figure}[htbp]
\includegraphics[width=3.5in]{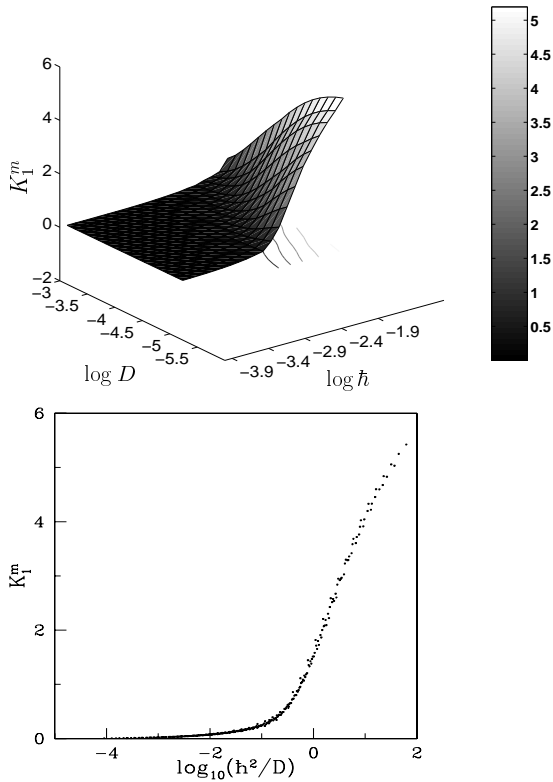}
\caption{\label{fig:NewMeas} Top: Maximal Kullback-Liebler 
  distance $K_1^m$ as a function of $\hbar$ and $D$, for the Quantum 
  Cat Map. Note that small values reflect strong similarity between
  classical and quantum evolutions. Bottom: Same data plotted in terms 
  of a composite parameter reflecting scaling behavior.}
\end{figure}

We illustrate the technique by considering a simple but extensively
studied system, the noisy quantum Cat Map~\cite{Josh,koslovsky,pb}. 
The classical limit displays extreme (uniformly hyperbolic) chaos,
and as such the system should be a member of a distinct universality 
class. The uniform hyperbolicity also precludes any dependence on 
initial conditions. The dynamics derive from the kicked 
oscillator Hamiltonian~\cite{ford} 
\begin{equation}
H=p^2/2\mu +\epsilon q^2/2\sum_{s=-\infty}^{\infty}\delta(s-t/T).
\label{fh}
\end{equation}
restricted to the torus $0\leq q < a$, $0\leq p < b$, with
the parameter constraints  $Tb/\mu a =1$ and $-\epsilon Ta/b=1$. 
The chaos here results not from the non-linearity of the Hamiltonian but 
from the choice of (re-injected) boundary conditions. As such, the general 
equation Eq.~(\ref{wigner}) does not apply. However, the first 
quantum correction to the classical propagator for this system 
(for the the Fourier-transformed distribution) is of order 
$\hbar k$ for the Fourier mode $k$~\cite{Josh}. The quantum-classical 
distance for this system then behaves as $\hbar\chi$, implying
that~\cite{fn2} $\zeta = \hbar^2\chi^2=\hbar^2\Lambda D^{-1}$. The 
top panel of Fig.~\ref{fig:NewMeas} shows $K^m_1$ as a function 
of $\hbar,D$. It is clear the distance behaves as expected. 
For example, as $\hbar$ is increased, larger $D$ values are needed for
the quantum and classical distributions to coincide. The lower panel 
shows the same data, plotted as a function of the single composite
variable $\zeta = \hbar^2/D$. The reduction of the surface in the upper panel 
to a single function of $\zeta$ demonstrates the scaling relationship between 
$\hbar,D$.  The accuracy of this scaling is reflected in the lack of any 
discernible spread around the curve. Remarkably, the scaling extends 
over many orders of magnitude in both parameters $\hbar,D$ and a 
considerable range in $K^m_1$. 

\begin{figure}[htbp]
\includegraphics[width=3.5in]{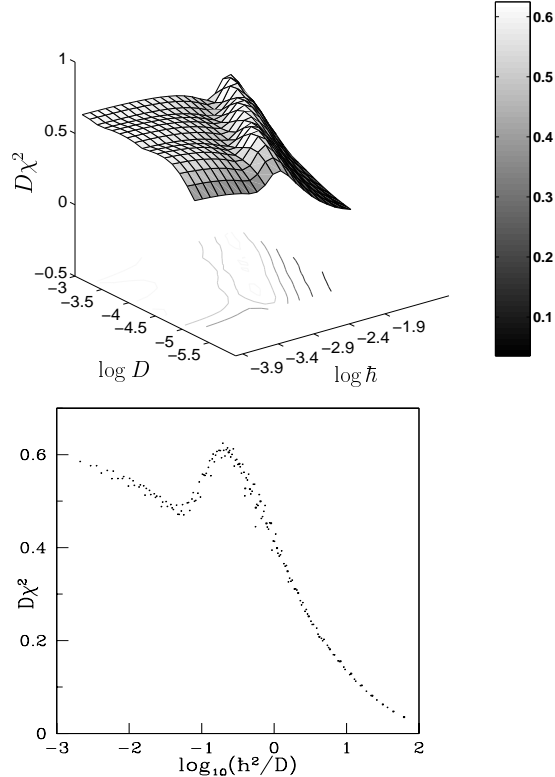}
\caption{Top: This measure reflects the generation of fine-scale
  structure in the dynamics with larger values corresponding to
  classical dynamics. Bottom: Same data plotted in terms of a
  composite parameter. Note the same scaling as in
  Fig~\ref{fig:NewMeas} and the coincidence of the transition region.}
\label{fig:OldMeas}
\end{figure}

The functional dependence of $K^m_1$ on $\zeta$ shows a number of 
distinctive features.
(i) $K^m_1$ is monotonic in $\zeta$, although as we argue below, 
there is no general reason to expect this. (ii) The quantum-classical 
distance is nonlinear in $\zeta$, with $K^m_1(\zeta)$ initially growing 
slowly as a function of $\zeta$, followed by a rapid transition at 
$\ln(\zeta)\approx 0$ or $\zeta\approx 1$.  This boundary is 
consistent with previous results~\cite{koslovsky,pb,ap}. (iii) The 
distance $K_1$ is bounded due to the noise, and we see the expected 
saturation for higher values of $\zeta$. (iv) There appear to be 
distinct regimes corresponding to small (for $\zeta <1$) and large
(for $\zeta >1$) quantum-classical distance. This last behavior is 
arguably generic as, in chaotic systems, a classical distribution develops 
fine-scaled structure very quickly ($\chi^2$ grows rapidly), increasing its
entropy production rate as well as its sensitivity to external noise. 
For this class of systems, in the first regime ($\zeta <1$), a quantum 
distribution initially remains close to the classical and will also 
increase its entropy production rate, and consequently the rate at which 
it becomes a mixed state. Hence any quantum effects that develop 
will be suppressed by the noise and the quantum-classical distance 
will remain small for all times. In this regime, the environment 
{\em minimizes} the quantum-classical difference. In the second regime 
(for $\zeta >1$), the quantum distribution does not initially follow the 
classical distribution to finer scales, and does not become sensitive to 
noise. It thus remains far from classical even as the noise alters the 
classical system. Here, the environment {\em exaggerates} the differences 
between quantum and classical probability dynamics. As such, 
$\zeta\approx 1$ may be viewed as a `quantum-classical boundary', with 
qualitatively different behavior on either side of it.

The general arguments above imply that similar scaling should be 
visible in all appropriately constructed measures of the quantum-classical
distance. In Fig.~\ref{fig:OldMeas} we show results for an alternate measure 
$D\chi^2$, which is related to the spreading of structure to finer 
scales. Unlike $K^m_1$ which compares classical and quantum evolution, 
this second measure is strictly quantum mechanical.  The supremum value in 
time of $D\chi^2$ ($\equiv D\chi^2_m$) is considered with varying $\hbar,D$ 
and for the same time-scales as before (classically, we would get a 
constant~\cite{physicaD}). Again, the precision and range of the scaling is 
remarkable. The qualitative conclusions are exactly the same as for 
$K^1_m(\zeta)$, with a similar rapid transition between large and small 
values of $D\chi^2_m$, happening again at $\zeta \approx 1$. That is, for 
small $\zeta$, the distribution is very sensitive to noise, changing 
rapidly as a function of $\zeta$ to low sensitivity.  However, this curve has a 
distinctive dip near $\zeta\approx 1$, such that the peak is at finite $\zeta$. 
This has been seen previously~\cite{97_1}, and can be understood by the fact 
that for near-classical quantum dynamics, the quantum follows the 
classical distribution but carries interference fringes on top of 
the classical structure. As such, the quantum distribution can be 
more sensitive to noise than the classical counterpart. 
In particular, as above, $\rho_W \approx \rho_c + a\hbar\chi\rho$ 
where $a$ is some constant. 
Similarly, the quantum and classical $\chi^2$ are related as
$\chi_q^2 \approx \chi_c^2 + a\hbar\chi^3$ so that to zeroth
order $\chi_{q0}^2 = \chi_c^2$, where the subscript on $\chi_q$ indicates
the order. To first order, we substitute the zeroth order expression
for $\chi^3$ to get $\chi_{q1}^2 \approx \chi_c^2 + a\hbar\chi_c^3$.
Iterating this procedure, to second order we will get terms like~\cite{fn3}
$\chi_{q2}^2 \approx \chi_c^2 + a\hbar\chi_c^3
(1 + a\hbar\chi_c)^{\frac{3}{2}}$. For small $a\hbar$ this becomes
\begin{eqnarray}
\chi_q^2 &\approx& \chi_c^2( 1 + a\;\hbar\chi_c + 
\frac{3}{2}a^2\;\hbar^2\chi_c^2 + \frac{3}{8}a^3\;\hbar^3\chi_c^3 +
\ldots)\nonumber \\
&\approx& \chi_c^2( 1 + a'\;\zeta^{\frac{1}{2}} + 
b\;\zeta + c\;\zeta^{\frac{3}{2}} 
+ \ldots)\label{chi-zeta}
\end{eqnarray}
where the constants $a',b,c$ absorb all other constants and we have
substituted $\hbar^2\chi^2 = \zeta$. The initial 
effect of quantum dynamics is to reduce the value of $\chi^2$ and 
hence $a'$ (and consequently $c$) must be negative valued constants, 
while $b$ is positive. For appropriate values of $a',b,c$, 
Eq.~(\ref{chi-zeta}) can indeed account for the shape of the curve 
seen in Fig.~(\ref{fig:OldMeas}). Therefore, all measures of 
quantum-classical distance need not depend monotonically on the system 
parameters. However, the particular dependence shown is almost 
definitely not generic since it depends on the relevant constants 
being of the appropriate ratios.

These results provide definitive evidence of parameter scaling 
in QCT for chaotic systems, which may be used 
to clearly identify different regimes of quantum-classical 
correspondence. As such, these are the first steps towards 
identifying and using composite parameters in studying universal 
behavior in the quantum-classical transition for small $\zeta$
(the near-classical regime). The smoothness and breadth of the 
scaling results shown are likely to be a feature of the uniform 
hyperbolicity of systems like the Cat Map. Understanding how this 
is altered by less extreme dynamics is clearly the next step, 
both in terms of constructing $\zeta$ and well as exploring the 
dependences of computed measures on $\zeta$. In particular, a 
preliminary assessment of entirely different measures applied to 
the quantum Duffing problem indicates that similar scaling may 
exist there as well.
 
{\em Acknowledgement} - A.K.P. acknowledges with pleasure useful
comments from Doron Cohen and Ivan Deutsch. The work of B.S. was 
supported by the National Science Foundation grant \#0099431 and 
a grant from the City University of New York PSC-CUNY Research 
Award Program.


\begin{thebibliography}{99}
\bibitem{fn1} A dependence of the QCT on $\zeta$ implies such a scaling 
dependence exists in the parameter $\zeta^r$ as well, for arbitrary $r$. 
We may choose one coefficient (e.g., setting $\alpha=2$) to determine the 
other two coefficients.
\bibitem{fn2} We set $\zeta =\hbar^\alpha\Lambda^\beta D^\gamma$, and 
choose $\alpha =2$ as discussed in \cite{fn1} above. Our figure
has a logarithmic scale for $\zeta$, whence this choice affects only 
the aspect ratio of the figure.
\bibitem{fn3} In general, the expansion for $\chi_q$ around 
$\chi_c$ in terms of $\zeta$ includes terms in $\zeta$ arising
from the higher-order terms in the relationship between 
$\rho_W$ and $\rho_c$. For the argument that the relationship 
between $\chi_q$ and $\chi_c$ is nonlinear in $\zeta$, these 
may be neglected since they only add to the nonlinearity.
\end{thebibliography}
\end{document}